\documentclass[aps,prb,10pt,twocolumn,superscriptaddress,floatfix]{revtex4-1}
\usepackage{chngcntr}
\usepackage{epstopdf}
\usepackage{graphicx}
\usepackage{dcolumn}
\usepackage{bm}
\usepackage{bbm,bbold}
\usepackage{color}
\usepackage{xcolor, soul}
\sethlcolor{green}
\usepackage{amsfonts}
\usepackage{amsmath}
\usepackage{mdframed}
\usepackage[normalem]{ulem}
\usepackage{mathrsfs}   
\usepackage[percent]{overpic}
\usepackage[colorlinks=true,citecolor=blue]{hyperref}
\hypersetup{colorlinks=true,citecolor=blue,linkcolor=blue,urlcolor=blue}

\begin{document}
\title{High-power collective charging of a solid-state quantum battery}
\author{Dario Ferraro}
\email{Dario.Ferraro@iit.it}
\affiliation{Istituto Italiano di Tecnologia, Graphene Labs, Via Morego 30, I-16163 Genova, Italy}
\author{Michele Campisi}
\affiliation{Istituto Italiano di Tecnologia, Graphene Labs, Via Morego 30, I-16163 Genova, Italy}
\author{Gian Marcello Andolina}
\affiliation{NEST, Scuola Normale Superiore, I-56126 Pisa, Italy}
\affiliation{Istituto Italiano di Tecnologia, Graphene Labs, Via Morego 30, I-16163 Genova, Italy}
\author{Vittorio Pellegrini}
\affiliation{Istituto Italiano di Tecnologia, Graphene Labs, Via Morego 30, I-16163 Genova, Italy}
\author{Marco Polini}
\affiliation{Istituto Italiano di Tecnologia, Graphene Labs, Via Morego 30, I-16163 Genova, Italy}
\begin{abstract}
Quantum information theorems state that it is possible to exploit collective quantum resources to greatly enhance the charging power of quantum batteries (QBs) made of many identical elementary units. We here present and solve a model of a QB that can be engineered in solid-state architectures. It consists of $N$ two-level systems coupled to a single photonic mode in a cavity. We contrast this collective model (``Dicke QB''), whereby entanglement is genuinely created by the common photonic mode, to the one in which each two-level system is coupled to its own separate cavity mode (``Rabi QB''). By employing exact diagonalization, we demonstrate the emergence of a quantum advantage in the charging power of Dicke QBs, which scales like $\sqrt{N}$ for $N\gg 1$.
\end{abstract}

\maketitle

\noindent{\it Introduction.---}In the last few decades, batteries~\cite{Vincent_and_Scrosati,Dell_and_Rand} have been the driving force behind the revolution in personal electronics and are steadily gaining tremendous importance also in the automotive sector~\cite{Electrical_Cars}. Currently, there is also an ever increasing demand on energy storage systems able to manage large power densities~\cite{luo_appliedenergy_2015}, an issue that has been so far partially addressed by the use of supercapacitors~\cite{wang_chemsocrev_2012,badwal_frontchem_2014}. Batteries and supercapacitors essentially operate on the basis of extremely robust electrochemical principles that have been developed between the Eighteenth and  Nineteenth centuries~\cite{Vincent_and_Scrosati,Dell_and_Rand}. While it is pivotal to continue research on advanced materials (see, e.g., Ref.~\onlinecite{bonaccorso_science_2015}) to optimize the performance of available energy storage devices, it seems timely and very natural to ask ourselves whether it is useful to transcend conventional electrochemistry to create an entirely new class of powerful batteries. 

Quantum phenomena, such as phase coherence and entanglement, constitute remarkable resources that, when properly manipulated and engineered, may enable superior performance of technological devices of various sorts. The prime example is quantum computing performed with quantum bits (realized, e.g., with superconducting circuitry~\cite{blais_pra_2004,devoret_science_2013}) as compared to classical computing performed with classical bits~\cite{DiVincenzo95}. While in quantum computing quantum phenomena are employed to achieve efficient manipulation and processing of information, an emerging theoretical research activity is currently focused on utilizing genuine quantum resources to achieve superior performances in the manipulation and processing of energy~\cite{Vinjanampathy16}. Specifically, whether and how quantum correlations may be harnessed to achieve high thermodynamic efficiency in the transduction of heat into work in quantum-mechanical thermal machines is something that is currently being actively investigated~\cite{Uzdin15, Campisi16, Karimi16, Marchegiani16, Hardal15,bera_arxiv_2016,bera_arxiv_2017,perarnau_arxiv_2017, Karimi17QST2}.

\begin{figure}[t]
\centering
\begin{overpic}[width=8.2cm]{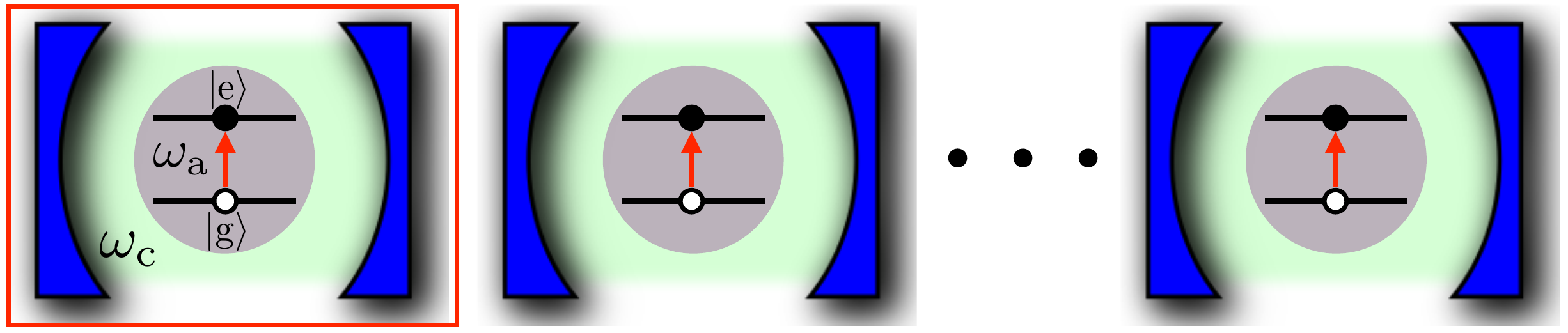}
\put(-5,16){(a)}
\end{overpic}
\begin{overpic}[width=8.2cm]{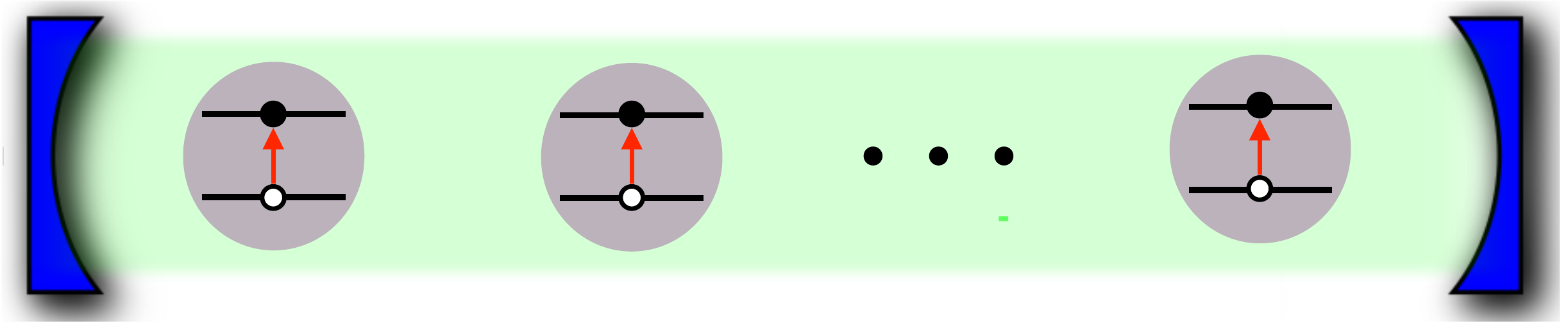}
\put(-5,16){(b)}
\end{overpic}
\begin{overpic}[width=8.17cm]{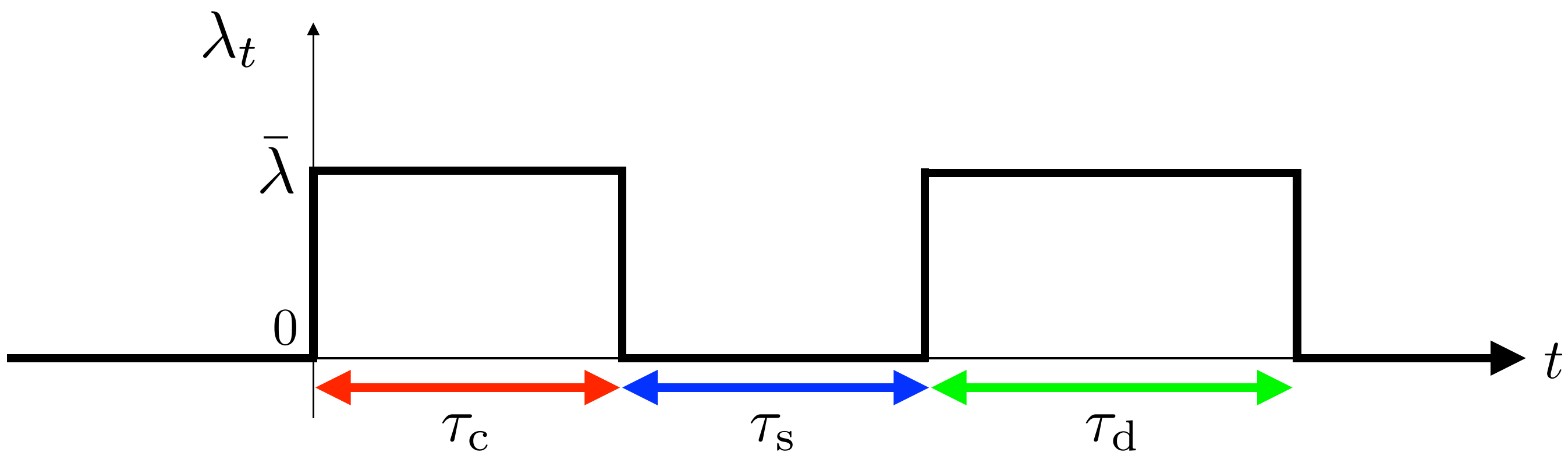}
\put(-5,26){(c)}
\put(28,9){i)}
\put(47,9){ii)}
\put(69,9){iii)}
\end{overpic}
\caption{(Color online) a) An array of identical Rabi quantum batteries operating in parallel. The elementary building block, enclosed in a red box, consists of a two-level system with an energy separation $\hbar \omega_{\rm a}$ between the ground $|{\rm g}\rangle$ and excited state $|{\rm e}\rangle$. Each two-level system is coupled to a separate cavity (blue) hosting a single photonic mode. The red arrow indicates a particle-hole transition induced by the photon field. b) A Dicke quantum battery, where the same array of two-level systems is now embedded into a single cavity and interacting with a common photonic mode. c) Time evolution of the dimensionless coupling constant $\lambda_{t}$ introduced in Eq.~(\ref{eq:Dicke}). i) After initializing the system in the state $|\psi^{(N)}(0)\rangle$ defined in Eq.~(\ref{eq:initialstate}), one turns on the coupling $\lambda_{t}$ bringing it to a finite value $\bar{\lambda}$ and keeping it on for a time $\tau_{\rm c}$ in order to charge the array (charging phase, red arrow). ii) The coupling is then switched off for a time $\tau_{\rm s}$ (storage phase, blue arrow). iii) Finally, the coupling is again turned on and set at the value $\bar{\lambda}$ for a time $\tau_{\rm d}$ to discharge the array (discharging phase, green arrow).\label{fig1}}
\end{figure}
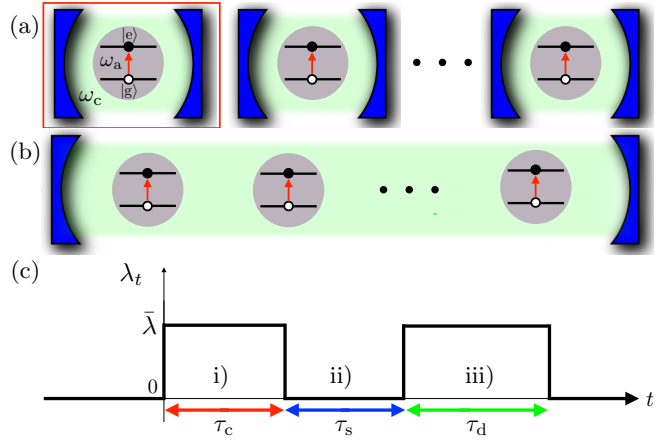

Given this context, we are naturally led to consider whether  quantum resources such as entanglement may be usefully employed to improve the performance (e.g.~by speeding up the charging time) of ``quantum batteries'' (QBs). To this end we consider a quantum system---see red box in Fig.~\ref{fig1}(a)---having a discrete energy spectrum, which can be kept well isolated from its environment so as to hold its energy for a sufficiently long time relative to the intended use. Many of such systems can be considered together, making a QB. In Figs.~\ref{fig1}(a) and~(b) we see two  examples. In panel (a), each quantum system---in this case a two-level system (TLS)---is coupled to a separate cavity, each hosting a single photonic mode. In panel (b), an ensemble of many TLSs is embedded in a single cavity and interacting with a common photonic mode. Charging of a QB requires a protocol of ``interaction'' of the QB itself with some external body or field (the ``energy source'', namely the cavity field in our example), which raises its energy over a time span that is much shorter than the QB lifetime. 

Early pioneering works~\cite{Alicki13, Hovhannisyan13, Binder15, Campaioli17} have considered a special subclass of charging protocols, namely those that can be described by unitary quantum gates acting on arrays of QBs. The main finding of these works is that addressing $N$ QBs at once, by means of a {\it global entangling operation}, can result in a speed-up of the averaged charging power (stored energy over charging time) as compared to charging them individually, in a parallel fashion. As noted in Ref.~\onlinecite{Campaioli17}, however, such global entangling operations involve highly non-local interactions, which may be difficult to realize in practice. 

In this work, we propose a practical architecture for a QB constituted by an array of $N$ TLSs (see Fig.~\ref{fig1}). We fully relax the constraint on the unitary evolution of the QB employed in previous theoretical works. Such unitary evolution regime occurs only when the dynamics of the energy source is very slow compared to the QB dynamics (i.e.~in the Born-Oppenheimer limit). This, although certainly interesting, is motivated more by mathematical convenience than adherence to reality. Here, we consider the more realistic situation in which no time scale separation exists between the QB and energy source subsystems. Accordingly, we treat the system ``QB+energy source'' in a fully quantum mechanical fashion, which generally results in a non-unitary reduced dynamics of the QB alone. In the proposed architecture, the non-local interaction among the $N$ TLSs is achieved by coupling all of them to the very same quantum energy source, which effectively results in a highly non-local entangling interaction. 

More specifically, our analysis relies on modelling the array of TLSs, entangled by a common quantized electromagnetic energy source, through the Dicke model~\cite{Dicke54}, see Fig.~\ref{fig1}(b). Interestingly, we find a quantum collective enhancement of the charging power of a factor $\sqrt{N}$, independent of the strength of the TLS-cavity coupling.

The solid-state Dicke QBs proposed here can be realized, e.g., by utilizing superconducting qubits~\cite{Wallraff04,Clarke08,Fink09,Xiang13} or nanofabricated semiconductor quantum dots~\cite{Ben04,Stockklauser17,singha_science_2011,Kung12,Basset13,Rossler15,Hofmann16,hensgens_arxiv_2017}, the former having relaxation and coherence times of the order of microseconds, with ultrafast (e.g.~nanosecond) charging time scales enabled by $\pi$ pulses. Another intriguing realization could rely on deterministically placing individual core-shell colloidal quantum dots~\cite{comin_chemsocrev_2014,detrizio_chemrev_2016} in an optical microcavity. Such systems may enable to run the charging protocol via visible photons, thus facilitating the experimental realization and increasing the stored energy density.

\noindent{\it Model and charging/discharging protocol.---}We consider the charging process of $N$ TLSs prepared in their ground state $|{\rm g}\rangle$, via coupling to a single cavity mode residing in the $N$ photon Fock state~\cite{Fock_state_footnote} $|N\rangle$. The initial state of the total system therefore reads:
\begin{equation}\label{eq:initialstate}
|\psi^{(N)}(0)\rangle=|N\rangle \otimes\underbrace{|{\rm g}, {\rm g}, \dots, {\rm g}\rangle}_\text{$N$}~.
\end{equation}

We model the quantum dynamics of the $N$ TLSs coupled to a single cavity mode via the following time-dependent Dicke Hamiltonian~\cite{Dicke54}:
\begin{align}\label{eq:Dicke}
\hat{\mathcal{H}}^{(N)}_{\lambda_t}=& \hbar \omega_{\rm c } \hat{a}^{\dagger} \hat{a}+ \omega_{\rm a} \hat{J}_{z} + 2 \omega_{\rm c} \lambda_t  \hat{J}_x \left(\hat{a}^{\dagger}+\hat{a}\right)~. 
\end{align}
Here, $\hat{a}$ ($\hat{a}^{\dagger}$) annihilates (creates) a cavity photon with frequency $\omega_{\rm c}$ and $\hat{J}_{\alpha}= (\hbar/2) \sum_i^N \hat{\sigma}_i^{\alpha}$ with $\alpha=x,y,z$ are the components of a collective spin operator expressed in terms of the Pauli operators $\hat{\sigma}^{\alpha}_i$ of the $i$-th TLS.
The quantity $\hbar \omega_{\rm a}$ is the energy splitting between the ground $|{\rm g}\rangle$ and excited state $|{\rm e}\rangle$ of each TLS. Below, we focus on the resonant regime, $\omega_{\rm a}  = \omega_{\rm c}$. The strength of the TLS-cavity coupling is given by the dimensionless parameter $\lambda_t$, whose explicit dependence on time $t$ specifies the charging/discharging protocol. 
For the sake of definiteness, we consider the protocol sketched in Fig.~\ref{fig1}(c). 
i) The interaction between the TLSs and the cavity is turned on at time $t=0^{+}$, $\lambda_{0^{+}}=\bar{\lambda}$, and kept it at this value for $0< t \leq \tau_{\rm c}$. During this charging step, energy transfer occurs from the cavity to the array of TLSs. ii) The interaction is then turned off at time $\tau^{+}_{\rm c}$, i.e.~$\lambda_{\tau^{+}_{\rm c}}=0$, and kept it off for $\tau_{\rm c} < t \leq \tau_{\rm c}+ \tau_{\rm s}$. During this storage step, the TLSs are assumed to be isolated from the environment, and hence keep their energy. Finally, iii) the interaction is again turned on for a time $\tau_{\rm d}$, $\lambda_{t}=\bar{\lambda}$ for $\tau_{\rm c}+ \tau_{\rm s} < t \leq \tau_{\rm c}+ \tau_{\rm s}+ \tau_{\rm d}$. During this discharging step, energy is transferred from the TLSs to the cavity. An alternative charging/discharging protocol, which is fully feasible experimentally~\cite{Fink09,Hofheinz08}, may rely on a time-independent $\lambda_{t} \to g$ coupling  and a non-zero time-dependent $\Delta_{t} = \omega_{\rm a}(t) - \omega_{\rm c}$ detuning. This is discussed in Appendix~\ref{AppA}, where the equivalence of these two alternative protocols is shown.

\noindent{\it Parallel charging.---}We begin by considering the case in which charging occurs in a parallel fashion, see Fig.~\ref{fig1}(a). Namely, we consider $N$ copies of TLSs, each coupled to a distinct cavity. In the case of a single TLS, 
the Dicke Hamiltonian (\ref{eq:Dicke}) reduces to the Rabi Hamiltonian~\cite{Rabi36,Rabi37}.
The energy $E^{(\|)}_{\bar{\lambda}}(\tau_{\rm c})$ stored at time $\tau_{\rm c}$ in a parallel fashion by $N$ copies of such resonant (i.e.~$\omega_{\rm a} = \omega_{\rm c}$) Rabi QBs is $N$ times the energy $\epsilon_{\bar \lambda}(\tau_{\rm c})$ stored in a single Rabi QB:
\begin{align}\label{epsilon}
E^{(\|)}_{\bar \lambda}(\tau_{\rm c}) = N \epsilon_{\bar \lambda}(\tau_{\rm c}) \equiv \frac{N \hbar \omega_{\rm c}}{2} &\Big[\langle \psi^{(1)}_{\bar \lambda}(\tau_{\rm c})| \hat{\sigma}_{z} | \psi^{(1)}_{\bar \lambda} (\tau_{\rm c}) \rangle \nonumber \\
&-\langle \psi^{(1)}(0)| \hat{\sigma}_{z} | \psi^{(1)}(0) \rangle \Big]~,
\end{align}
with $\hat{\sigma}_{z}=|{\rm e}\rangle\langle {\rm e}| - |{\rm g}\rangle\langle {\rm g}|$.

The label $\bar{\lambda}$ in $E^{(\|)}_{\bar \lambda}(\tau_{\rm c})$ reminds us that the stored energy depends on $\bar{\lambda}$. The symbol
$| \psi^{(1)}_{\bar \lambda}(\tau_{\rm c})\rangle$ stands for the evolved initial state $|\psi^{(1)}(0)\rangle$, according to $\hat{\mathcal{H}}^{(1)}_{\bar{\lambda}}$ for a time $\tau_{\rm c}$, i.e.,~$|\psi^{(1)}_{\bar \lambda}(\tau_{\rm c})\rangle = e^{-i \hat{\mathcal{H}}^{(1)}_{\bar{\lambda}} \tau_{\rm c}/\hbar}|\psi^{(1)}(0)\rangle$.

We now introduce the maximum stored energy (i.e.~the ``capacity'') and the maximum charging power in the parallel-charging operation mode: $E^{(\parallel)}_{\bar \lambda}=\max_{\tau_{\rm c}} [E^{(\parallel)}_{\bar \lambda}(\tau_{\rm c})]$ and $P^{(\parallel)}_{\bar \lambda}=\max_{\tau_{\rm c}}[P^{(\parallel)}_{\bar \lambda}(\tau_{\rm c})]$, where the charging power~\cite{Binder15,Campaioli17} after a time $\tau_{\rm c}$ is defined as $P^{(\parallel)}_{\bar \lambda}(\tau_{\rm c}) \equiv E^{(\parallel)}_{\bar \lambda}(\tau_{\rm c})/\tau_{\rm c}$.

Both  $E^{(\parallel)}_{\bar \lambda}$ and $P^{(\parallel)}_{\bar \lambda}$ scale linearly with $N$ (yielding a constant energy and power per QB):
\begin{equation}\label{E_lambda0}
E^{(\parallel)}_{\bar \lambda}= \hbar \omega_{\rm c} N \mathcal{F}_{\rm E}(\bar \lambda) \propto N
\end{equation}
and
\begin{equation}\label{P_lambda0}
P^{(\parallel)}_{\bar \lambda}= \hbar \omega^2_{\rm c} N \mathcal{F}_{\rm P}(\bar \lambda) \propto N~,
\end{equation} 
where $\mathcal{F}_{\rm E}(\bar \lambda)$ and $\mathcal{F}_{\rm P}(\bar \lambda)$ are dimensionless functions of $\bar \lambda$, which can be calculated exactly~\cite{Braak11}. Their expression greatly simplifies in the weak-coupling $\bar \lambda\ll1$ limit, where the Rabi Hamiltonian can be approximated by the Jaynes-Cummings one~\cite{Jaynes63}. The stored energy takes the form $E^{(\|)}_{\bar \lambda\ll 1}(\tau_{\rm c}) \to  N \hbar \omega_{\rm c}\sin^{2}\left(\bar \lambda \omega_{\rm c}\tau_{\rm c}\right)$, and hence~\cite{Prefactors_1}  $\mathcal{F}_{\rm E}(\bar{\lambda}\ll 1) \to 1$ and $ \mathcal{F}_{\rm P}(\bar{\lambda}\ll 1) \to 0.724 \bar{\lambda}$. Since we are interested in the collective charging case and in scalings with $N$, we will not dwell upon deriving exact expressions for $\mathcal{F}_{\rm E}(\bar{\lambda})$ and $\mathcal{F}_{\rm P}(\bar{\lambda})$.

{\it Collective charging.---}We now investigate the maximum stored energy and maximum charging power when the $N$ TLSs are coupled to one and the same cavity---see Fig.~\ref{fig1}(b)---as described by the Dicke Hamiltonian (\ref{eq:Dicke}). The latter has a conserved quantity given by the so-called cooperation number~\cite{Tavis68, Tavis69} $\hat{J}^{2}= \sum_{\alpha=x,y,z}\hat{J}^{2}_{\alpha}$. A convenient basis set for representing the Hamiltonian (\ref{eq:Dicke}) is $|n,j,m\rangle$, where $n$ indicates the number of photons, $j(j+1)$ is the eigenvalue of $\hat{J}^{2}$, and $m$ denotes the eigenvalue of $\hat{J}_{z}$. With this notation, the initial state (\ref{eq:initialstate}) reads $|\psi^{(N)}(0)\rangle=|N,\frac{N}{2}, -\frac{N}{2} \rangle$
and the matrix elements of the Dicke Hamiltonian read~\cite{Magnani11}
\begin{eqnarray}\label{eq:matrix_Dicke}
&&\langle n', \frac{N}{2}, \frac{N}{2}-q'|\hat{\mathcal{H}}^{(N)}_{\lambda_t} |n,\frac{N}{2}, \frac{N}{2}-q\rangle=\\ \nonumber 
&&\hbar \omega_{\rm c}\Bigg\{\left(n+\frac{N}{2}-q\right)\delta_{n', n} \delta_{q', q}\\ \nonumber 
 &&+\lambda \left[f^{(1)}_{n, \frac{N}{2}, \frac{N}{2}-q}\delta_{n', n+1} \delta_{q', q+1} +f^{(2)}_{n, \frac{N}{2}, \frac{N}{2}-q}\delta_{n', n+1}\delta_{q', q-1} \right. \\ \nonumber
 &&+\left. f^{(3)}_{n, \frac{N}{2}, \frac{N}{2}-q}\delta_{n', n-1}\delta_{q', q+1} +f^{(4)}_{n, \frac{N}{2}, \frac{N}{2}-q}\delta_{n', n-1} \delta_{q', q-1} \right]\Bigg\}
\end{eqnarray}
with
\begin{eqnarray}
f^{(1)}_{k, j, m}&=& \sqrt{(k+1)\left[j(j+1)-m(m-1) \right]}~,\\
f^{(2)}_{k, j, m}&=& \sqrt{(k+1)\left[j(j+1)-m(m+1) \right]}~,\\ 
f^{(3)}_{k, j, m}&=& \sqrt{k\left[j(j+1)-m(m-1) \right]}~,\\ 
f^{(4)}_{k, j, m}&=& \sqrt{k\left[j(j+1)-m(m+1) \right]}~.
\end{eqnarray}
We remark that the number of photons is not conserved by the Dicke Hamiltonian nor it is bounded from above, hence taking, in principle, any integer value. Accordingly, the matrix (\ref{eq:matrix_Dicke}) is infinite-dimensional. In practice, we need to truncate it by introducing a cutoff $N_{\rm ph}>N$ given by the maximum number $N_{\rm ph}$ of photons. This is chosen in such a way that a larger value of it, $N^{\prime}_{\rm ph} > N_{\rm ph}$, would not produce any noticeable difference in the computed eigenvalues and eigenvectors. 
In the following, we show numerical results obtained from the exact diagonalization of the matrix (\ref{eq:matrix_Dicke}) for $N=2, \dots, 20$. We have checked that excellent numerical convergence is achieved by choosing $N_{\rm ph}  = 4 N$. (A linear scaling of $N_{\rm ph}$ with $N$ has also been found in Ref.~\onlinecite{Magnani11}.)

\begin{figure}[t]
\begin{center}
\includegraphics[width=8.6cm]{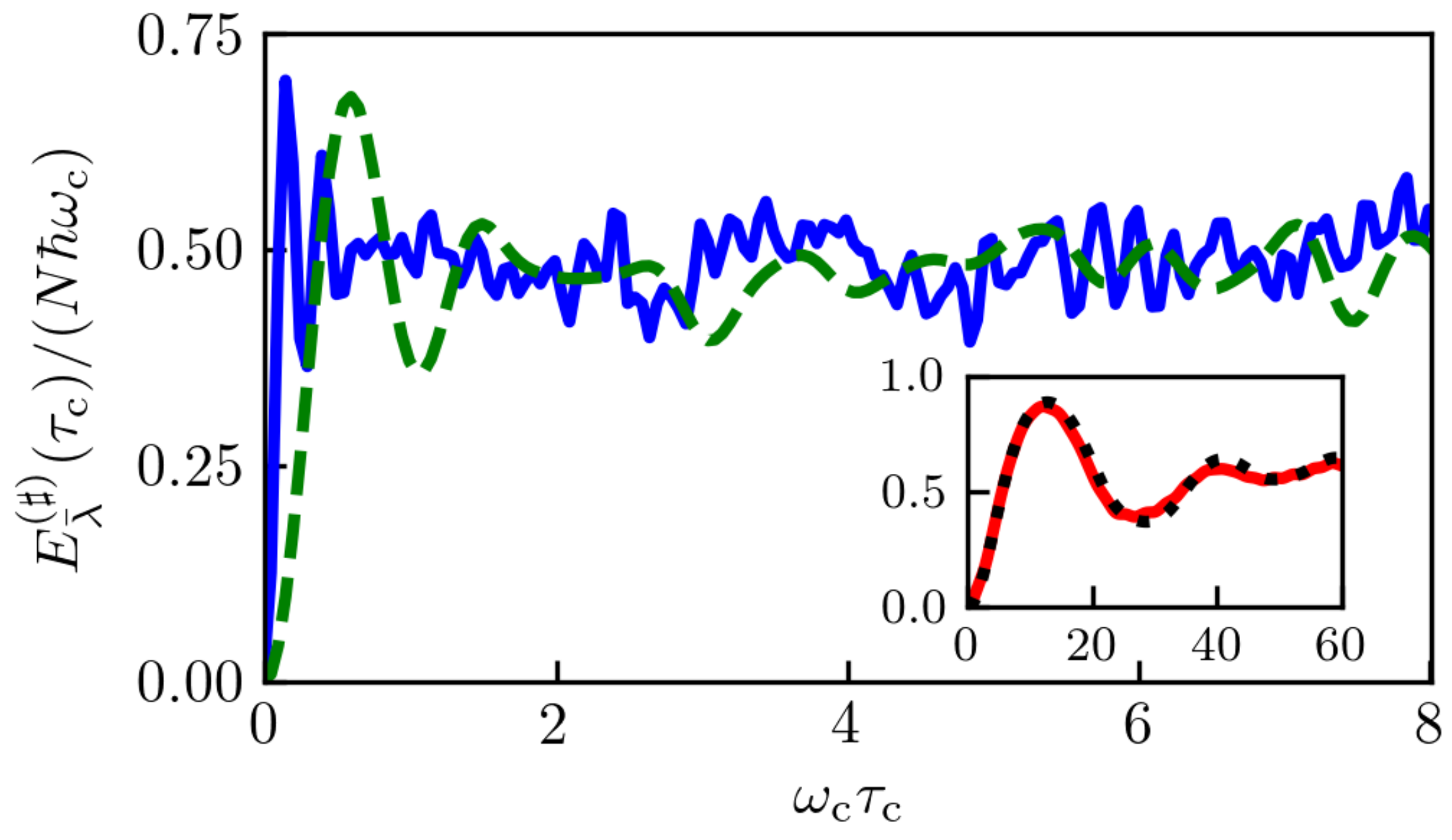}
\end{center}
\caption{(Color online) The dependence of the stored energy $E^{(\sharp)}_{\bar{\lambda}}(\tau_{\rm c})$ (in units of $N\hbar \omega_{\rm c}$) on 
$\tau_{\rm c}$ (in units of $1/\omega_{\rm c}$) for $\bar{\lambda}=0.5$ (green dashed line) and $\bar{\lambda}=2.0$ (blue solid line). Inset: same as in the main panel but for $\bar{\lambda}=0.05$ (red solid line), i.e.~in the weak-coupling limit, is compared to results for the Tavis-Cummings model (black dotted line). All data in this figure have been computed by setting $N=10$.\label{fig2}}
\end{figure}

The energy $E^{(\sharp)}_{\bar\lambda}(\tau_{\rm c})$ stored collectively at time $\tau_{\rm c}$ by the $N$ TLSs is given by
\begin{align}\label{E_stored}
E^{(\sharp)}_{\bar \lambda}(\tau_{\rm c}) = \omega_{\rm c}\Big[&\langle \psi^{(N)}_{\bar\lambda}(\tau_{\rm c})| \hat{J}_{z}|\psi^{(N)}_{\bar\lambda}(\tau_{\rm c})\rangle \nonumber\\
&-\langle \psi^{(N)}(0)| \hat{J}_{z}|\psi^{(N)}(0)\rangle\Big]~,
\end{align}
where $|\psi^{(N)}_{\bar\lambda}(\tau_{\rm c})\rangle=e^{-i \hat{\mathcal{H}}^{(N)}_{\bar{\lambda}}\tau_{\rm c}/\hbar}|\psi^{(N)}(0)\rangle$.
The dependence of $E^{(\sharp)}_{\bar\lambda}(\tau_{\rm c})$ on $\tau_{\rm c}$ is reported in Fig.~\ref{fig2} for a few values of $\bar{\lambda}$.
We observe smooth oscillations for $\bar{\lambda}\ll 1$ (red solid line in the inset of Fig.~\ref{fig2}), which are in full agreement with results obtained for the Tavis-Cummings model~\cite{Tavis68, Tavis69} (black dotted line in the inset of Fig. \ref{fig2}). In the latter, counter-rotating terms are absent, leading to the conservation of the number of excitations and the further constraint $n=q$ ($n'=q'$) in Eq. (\ref{eq:matrix_Dicke}). A more complicated oscillatory pattern showing beating appears for increasing $\bar{\lambda}$ (green dashed and blue solid lines in Fig.~\ref{fig2}).
\begin{figure}[ht] 
  \label{ fig3} 
  \begin{minipage}[c]{0.5\linewidth}
  \centering
 \begin{overpic}[width=4.45cm]{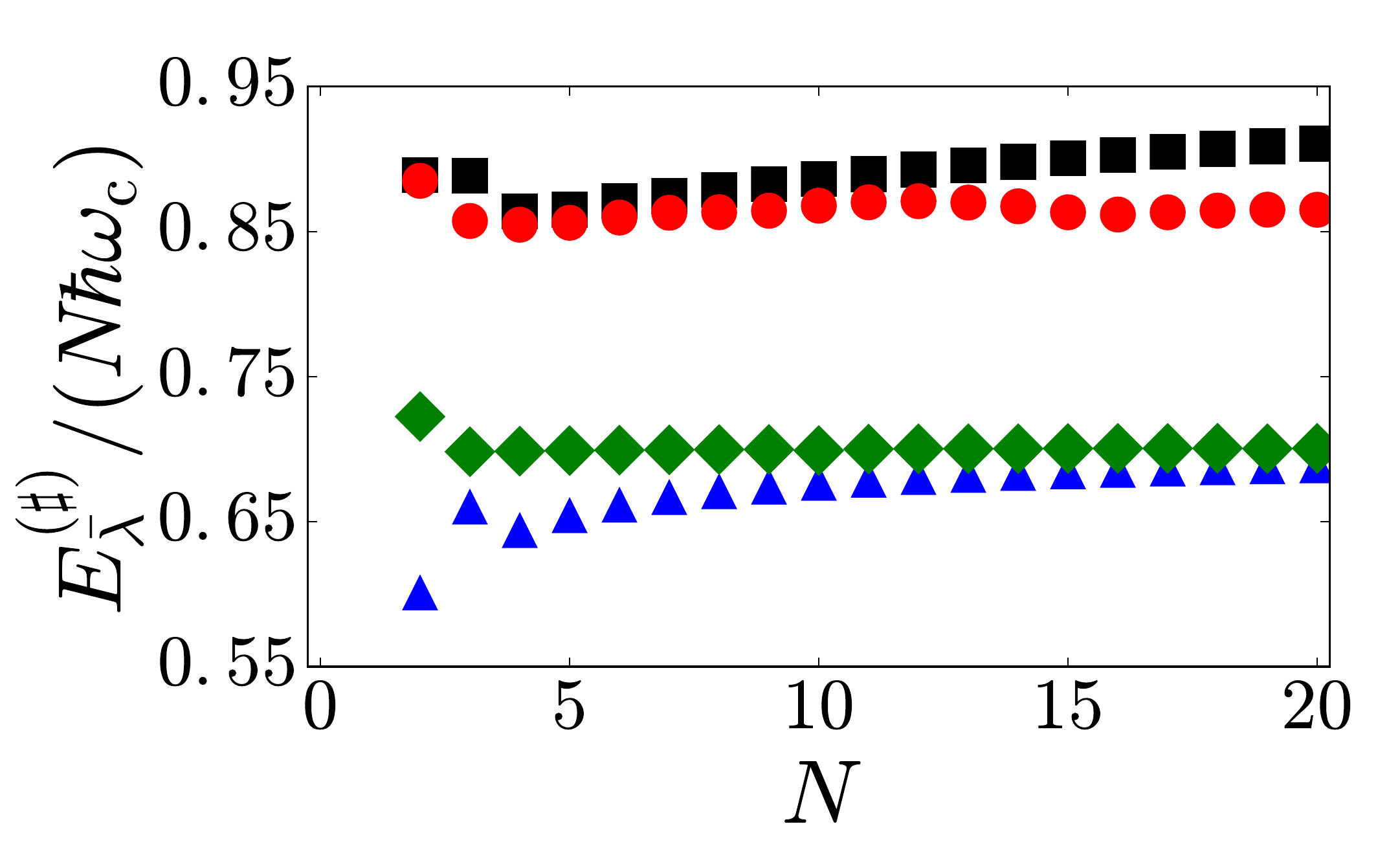}
 \put(1,64){(a)}
 \end{overpic}
  \vspace{4ex}
  \end{minipage}
  \begin{minipage}[c]{0.5\linewidth}
  \centering
 \begin{overpic}[width=4.45cm]{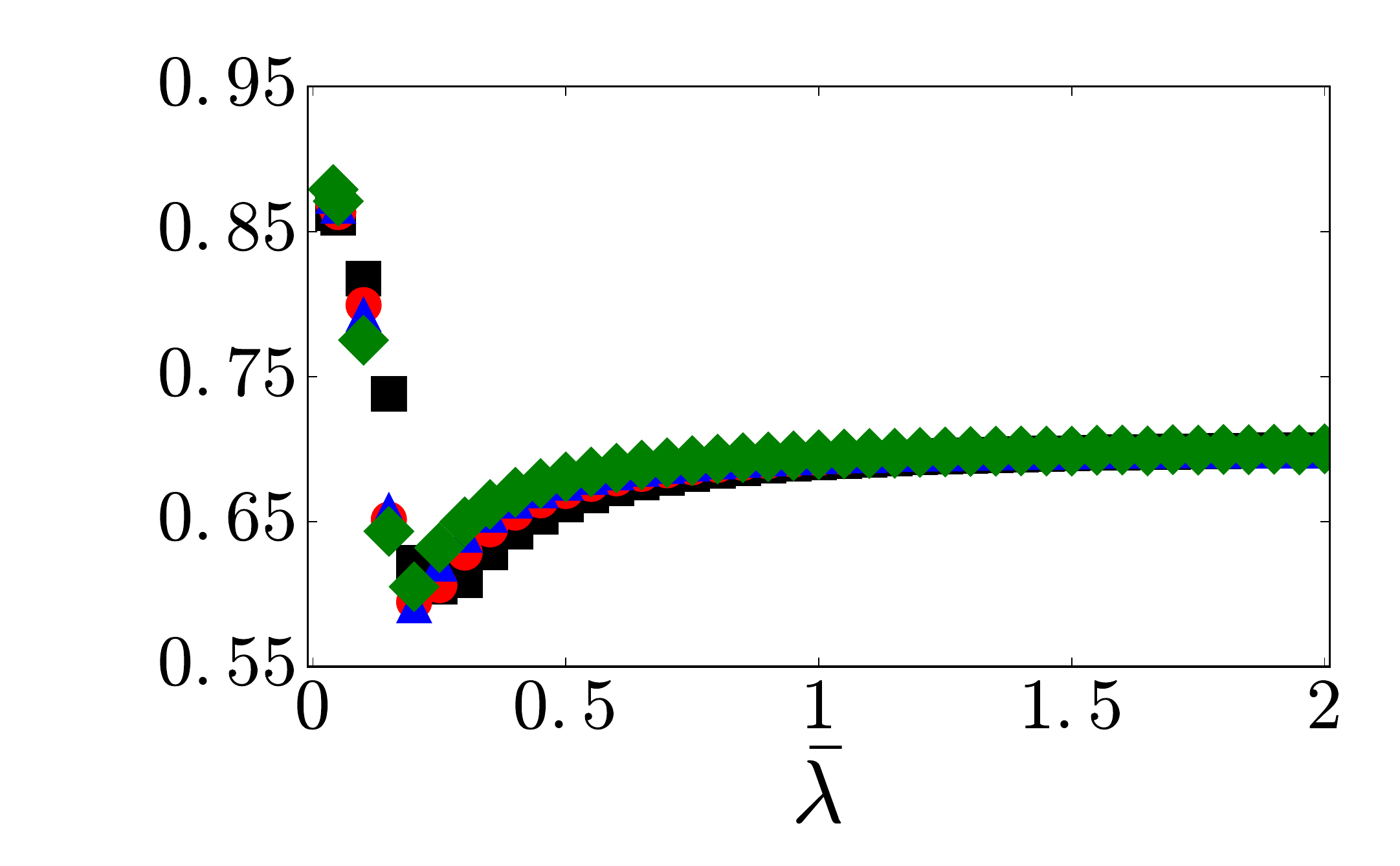}
 \put(9,64){(b)}
 \end{overpic}
  \vspace{4ex}
  \end{minipage} 
  \begin{minipage}[b]{0.5\linewidth}
  \centering
 \begin{overpic}[width=4.45cm]{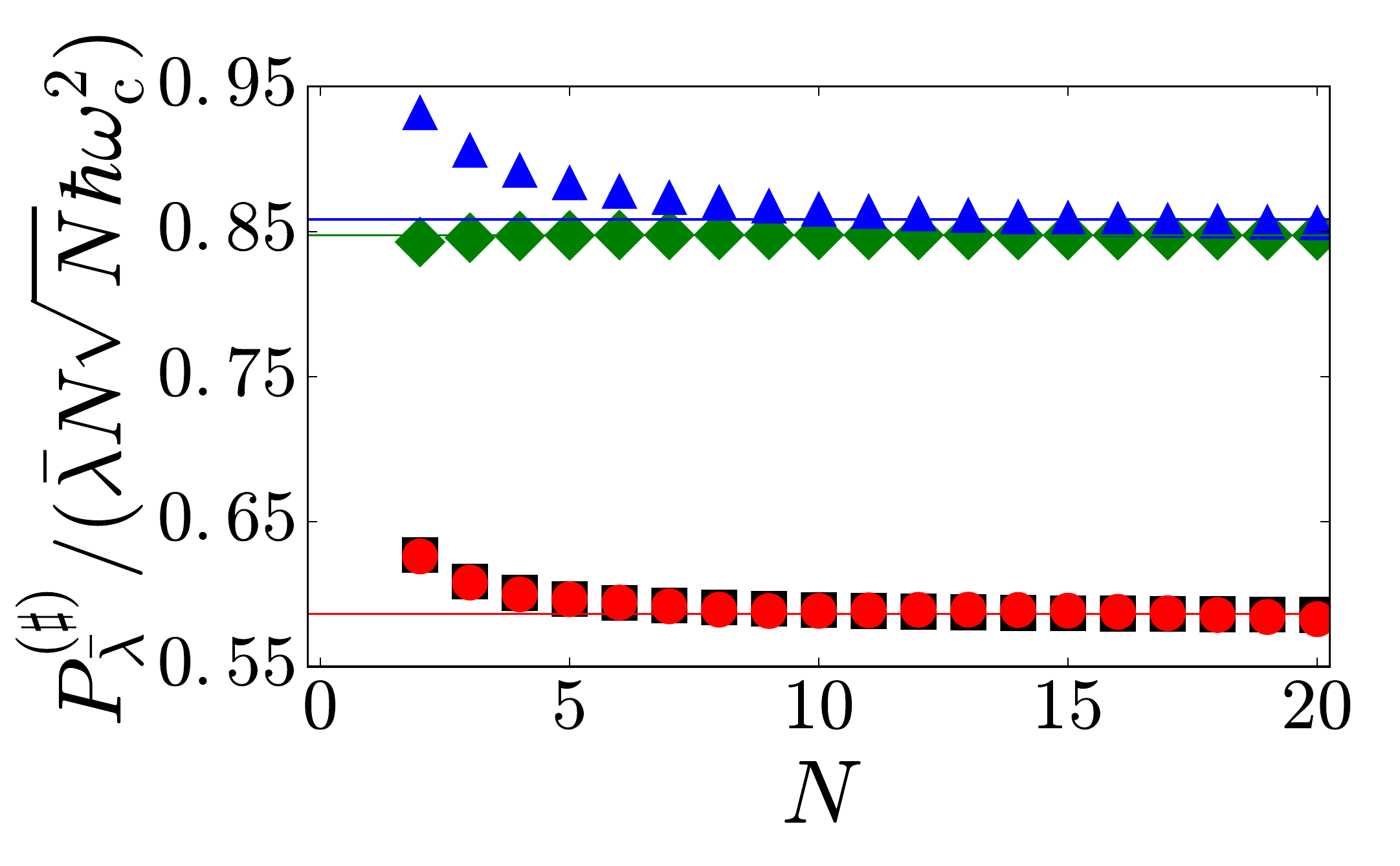}
 \put(1,64){(c)}
 \end{overpic}
 \vspace{4ex}
  \end{minipage}
  \begin{minipage}[b]{0.5\linewidth}
  \centering
 \begin{overpic}[width=4.45cm]{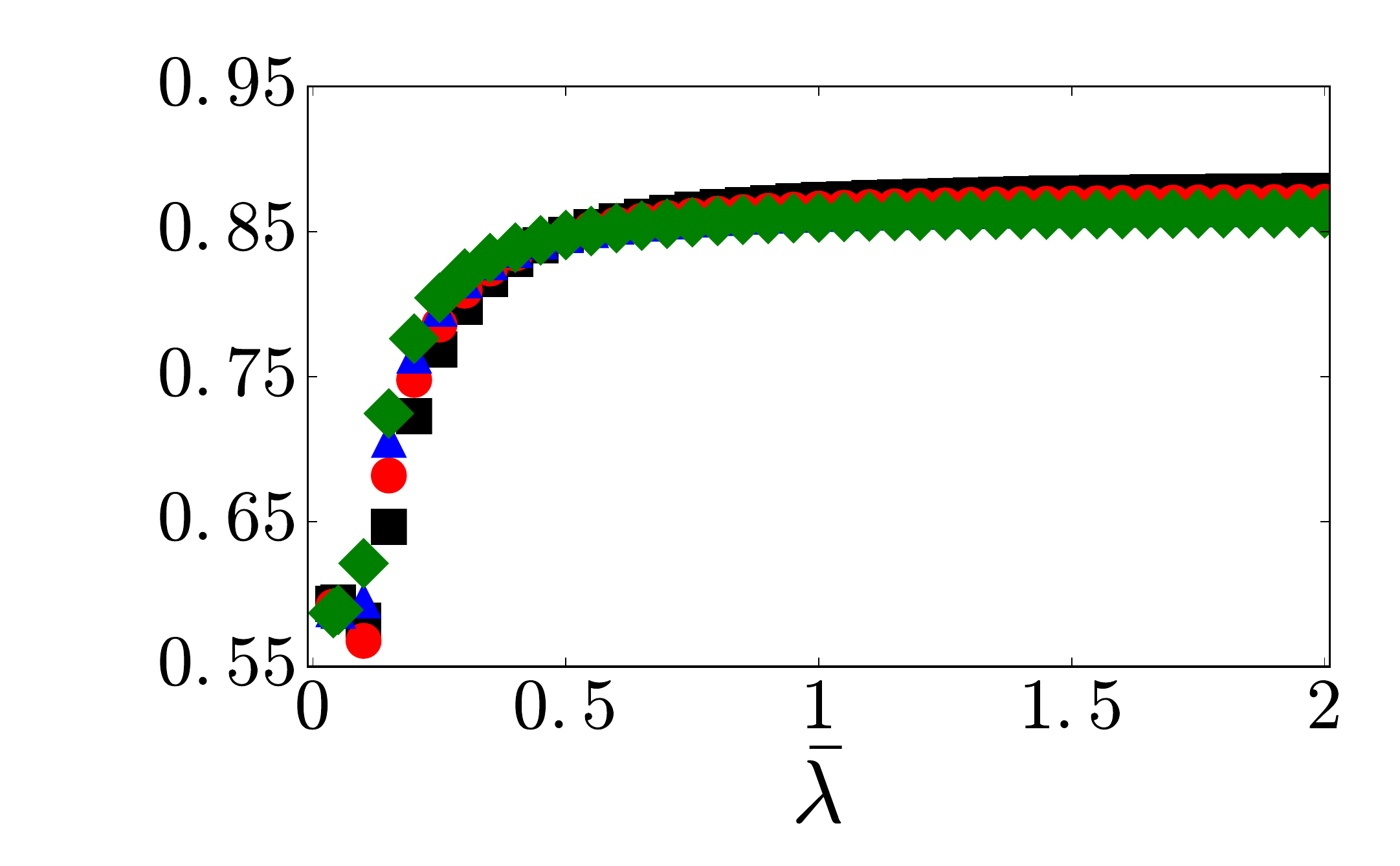}
 \put(9,64){(d)}
 \end{overpic}
    \vspace{4ex}
  \end{minipage} 
  \caption{(Color online) Panel (a) The maximum stored energy $E^{(\sharp)}_{\bar{\lambda}}$ (in units of $N\hbar \omega_{\rm c}$) is plotted as a function of $N$. Black squares denote the results for the Tavis-Cummings model at $\bar{\lambda}=0.05$. Results for Dicke QBs refer to $\bar{\lambda}=0.05$ (red circles), $\bar{\lambda}=0.5$ (blue triangles), and $\bar{\lambda}=2.0$ (green diamonds). Panel (b) Same as in panel (a) but as a function of $\bar{\lambda}$ for $N=6$ (black squares), $N=8$ (red circles), $N=10$ (blue triangles), and $N=12$ (green diamonds).  Panel (c) The maximum charging power $P^{(\sharp)}_{\bar{\lambda}}$ (in units of $\bar{\lambda} N\sqrt{N} \hbar \omega^{2}_{\rm c}$) is plotted as a function of $N$. Color coding and labeling is the same as in panel (a). The thin horizontal lines are best fits to the numerical results, indicating the asymptotic values of the maximum power at large $N$: $\lim_{N\gg 1}P^{(\sharp)}_{\bar{\lambda}}/(\bar{\lambda} N\sqrt{N} \hbar \omega^{2}_{\rm c}) = 0.586$ for $\bar \lambda =0.05$ (red), $0.858$ for $\bar\lambda =0.5$ (blue), and $0.847$ for $\bar \lambda =2$ (green).  Panel (d) Same as in panel (c) but as function of $\bar{\lambda}$ for $N=6$ (black squares), $N=8$ (red circles), $N=10$ (blue triangles), and $N=12$ (green diamonds).\label{fig3}}
\end{figure}

Figures~\ref{fig3}(a) and (c) show the maximum stored energy $E^{(\sharp)}_{\bar\lambda} \equiv \max_{\tau_{\rm c}}[E^{(\sharp)}_{\bar\lambda}(\tau_{\rm c})]$ and maximum charging power $P^{(\sharp)}_{\bar\lambda} \equiv \max_{\tau_{\rm c}}[E^{(\sharp)}_{\bar\lambda}(\tau_{\rm c})/\tau_{\rm c}]$ in the collective case, as functions of $N$, for various values of $\bar\lambda$. Note that the vertical axes of Figs.~\ref{fig3}(a) and (c) are rescaled by $N$ and $\bar{\lambda}N\sqrt{N}$, respectively. 
We clearly see that such rescaled quantities rapidly converge to a certain asymptotic value as $N$ increases. 
This implies that, for sufficiently large values of $N$, $E^{(\sharp)}_{\bar\lambda} $ and $P^{(\sharp)}_{\bar\lambda}$ reach  asymptotic values characterized by the following scaling laws
\begin{align}\label{eq:scaling_power_energy}
E^{(\sharp)}_{\bar\lambda} \propto  N 
\end{align}
and 
\begin{align}\label{eq:scaling_power_collective}
P^{(\sharp)}_{\bar\lambda} \propto  N^{3/2}~. 
\end{align}
The super-linear scaling of the maximum charging power in Eq.~(\ref{eq:scaling_power_collective}) constitutes direct evidence of a $\sqrt{N}$ quantum advantage associated to collective charging~\cite{Campaioli17} as compared to parallel charging, Eq.~(\ref{P_lambda0}). Such advantage is related to a scaling law of the time required to reach the maximum power, $\tau_{\rm c} \propto 1/\sqrt{N}$, and has a quantum mechanical origin. The quantum speed-up is indeed associated with the fact that, unlike in the case of the parallel evolution, the collective evolution proceeds through states characterized by quantum entanglement among the TLSs. Thus our Dicke QBs clearly display the powerful charging mechanism described in abstract terms in Refs.~\onlinecite{Binder15, Campaioli17}. Finally, Figs.~\ref{fig3}(b) and (d) illustrate the dependencies of the maximum stored energy and charging power of Dicke QBs on the coupling constant $\bar\lambda$, for various values of $N$. Plotting the same rescaled quantities, $E^{(\sharp)}_{\bar\lambda}/(N\hbar \omega_{\rm c})$ and $P^{(\sharp)}_{\bar\lambda}/({\bar \lambda}N\sqrt{N} \hbar \omega^2_{\rm c})$, versus the effective coupling parameter~\cite{Emary03} 
${\bar \Lambda} \equiv {\bar \lambda}\sqrt{N}$, one notices a collapse onto universal curves, as shown in Appendix~\ref{app:universal_scaling}. We remind the reader that the ground state of the Dicke model displays a superradiant quantum phase transition (SQPT) 
at ${\bar \Lambda}=1/2$. A feeble reminiscence of such SQPT is also seen in the maximum charging energy of Dicke QBs, 
as illustrated in Appendix~\ref{app:universal_scaling}. See also Ref.~\onlinecite{footnote:Asquared_term}.

\noindent{\it Storage and discharging.---}We now briefly comment upon storage and discharging phases of our Dicke QBs. We assume the storage time $\tau_{\rm s}$ is much shorter than any decoherence/relaxation time scale in a real solid-state implementation. 
Under this assumption, the Dicke QB retains its energy during the storage step.
In the parallel case, and when ${\bar \lambda} \ll 1$ (in which case the rotating wave approximation holds), independent of the duration of the storage time $\tau_{\rm s}$, the initial state (\ref{eq:initialstate}) is recovered at the end of the discharging phase if the condition $\tau_{\rm c}+\tau_{\rm d}=\pi/(\bar \lambda\omega_{\rm c})$ is met. In the collective case, as either $N$ or $\bar\lambda$ increases, such recoverability is lost. Accordingly, the smaller $\bar\lambda$ the higher $E^{(\sharp)}_{\bar\lambda}$, the higher the recoverability (not shown). This is a signature of energy injection incurred when turning the coupling on and off, namely that  $\delta E^{\rm on}_{\bar \lambda}=\langle \psi^{(N)}_{\bar\lambda}(0)| \hat{\mathcal{H}}^{(N)}_{\lambda_{0+}}- \hat{\mathcal{H}}^{(N)}_{\lambda_{0-}}|\psi^{(N)}_{\bar\lambda}(0)\rangle $ and $\delta E^{\rm off}_{\bar \lambda}=\langle \psi^{(N)}_{\bar\lambda}(\tau_{\rm c})| \hat{\mathcal{H}}^{(N)}_{\lambda_{\tau_{\rm c}+}}- \hat{\mathcal{H}}^{(N)}_{\lambda_{\tau_{\rm c}-}}|\psi^{(N)}_{\bar\lambda}(\tau_{\rm c})\rangle $ are generally non-vanishing.

\noindent{\it Experimental considerations.---}The TLS+cavity system may be realized with state-of-the-art solid-state technology, by using e.g.~superconducting qubits coupled to superconducting line resonators~\cite{Wallraff04,Clarke08,Fink09} or nanofabricated quantum dots (see e.g.~Refs.~\onlinecite{singha_science_2011,hensgens_arxiv_2017}) combined with superconductive microwave circuits~\cite{hensgens_arxiv_2017,Stockklauser17}, photonic crystals~\cite{Ben04}, or THz planar microcavities~\cite{zhang_naturephys_2016}. Concerning the typical values of the relevant physical parameters discussed in this work, the implementations of Rabi and Dicke Hamiltonians in such solid-state devices~\cite{Fink09, Stockklauser17, Hofheinz08} have a resonant frequency $\omega_{\rm c}\approx\omega_{\rm a}$ ranging from GHz to THz values and an individual interaction parameter $g_{0}=\bar{\lambda} \omega_{\rm c}$ typically taking values in the range $10$-$100~{\rm MHz}$. This leads to $\bar \lambda\approx 10^{-3}$-$10^{-2}$, fully justifies the rotating wave approximation discussed for the Rabi model. Moreover, the relevant time scales of relaxation and decoherence processes have to be compared with $g_{0}^{-1}$. In particular, one can introduce the decoherence rate $\Gamma_{\phi}$ and the electron relaxation rate $\Gamma_{\rm e}$~\cite{blais_pra_2004,Schoelkopf08}. The proposed charging/discharging protocol---together with all other possible quantum-computing implementations---is meaningful under the condition $\Gamma_{\phi}\lesssim\Gamma_{\rm e}< g_{0}$, which is satisfied in the experiments discussed in Refs.~\onlinecite{Fink09, Stockklauser17, Hofheinz08}. This condition is even further justified in the Dicke model where the global coupling scales as $g=g_{0}\sqrt{N}$. Recent experimental work has also demonstrated that the strong-coupling $\bar\lambda \approx 1$ limit can also be reached~\cite{yoshihara_naturephys_2017,langford_arxiv_2016,braumueller_arxiv_2016}. Colloidal quantum dots such as core-shell CdSe dots~\cite{comin_chemsocrev_2014,detrizio_chemrev_2016} may offer another possible solution for implementing Dicke QBs, bringing the resonant frequency to hundreds of THz. This could facilitate the coupling of the dots with the photonic (micro)cavity mode and also yield an improved stored energy density.

{\it Conclusions.---}We have introduced the concept of a ``Dicke quantum battery'', consisting of an array of entangled two-level systems. Our aim is to put on concrete and experimentally feasible grounds the intriguing abstract ideas previously presented in Refs.~\onlinecite{Binder15,Campaioli17}. The main physics is captured by the toy model in Eq.~(\ref{eq:Dicke}), which can in principle be engineered in a solid-state platform and displays collective powerful charging, Eq.~(\ref{eq:scaling_power_collective}) and Fig.~\ref{fig3}(c). In particular, the interaction of an array of two-level systems with a common quantized electromagnetic mode in a cavity automatically creates entanglement among the $N$ two-level systems. This is ultimately due to an effective long-range interaction between the two-level systems mediated by the cavity photons.  We observe an enhanced scaling of the maximum charging power, as envisaged in Refs.~\onlinecite{Binder15, Campaioli17}, with a $\sqrt{N}$-fold enhancement with respect to the parallel case, independent of the value of the coupling strength $\bar{\lambda}$---see Eq.~(\ref{eq:scaling_power_collective}). We further note an interesting trade-off between power and reversibility of the charging process. Highest values of the maximum power are achieved at strong coupling. These come, however, at the cost of a lower stored energy, accompanied by a decrease in the efficiency of energy transfer from the quantum batteries to the cavity in the discharging phase. On the other hand, at weak coupling, one finds larger values of the maximum stored energy and a higher efficiency of energy transfer in the discharging 
step, at the cost of lower values of the maximum power.

\noindent{\it Acknowledgments.---}We thank M. Lewestein, A. Riera, M. Nath Bera, V. Giovannetti, A. Mari, F. Pellegrino, and M. Paternostro for useful discussions. M.C. wishes to thank the COST action MP1209 ``Thermodynamics in the Quantum Regime" for support.

\appendix

\counterwithin{figure}{section}
\section{Considerations on an  alternative charging/discharging protocol} 
\label{AppA}
\begin{figure}[t]
\centering
\includegraphics[width=8.2cm]{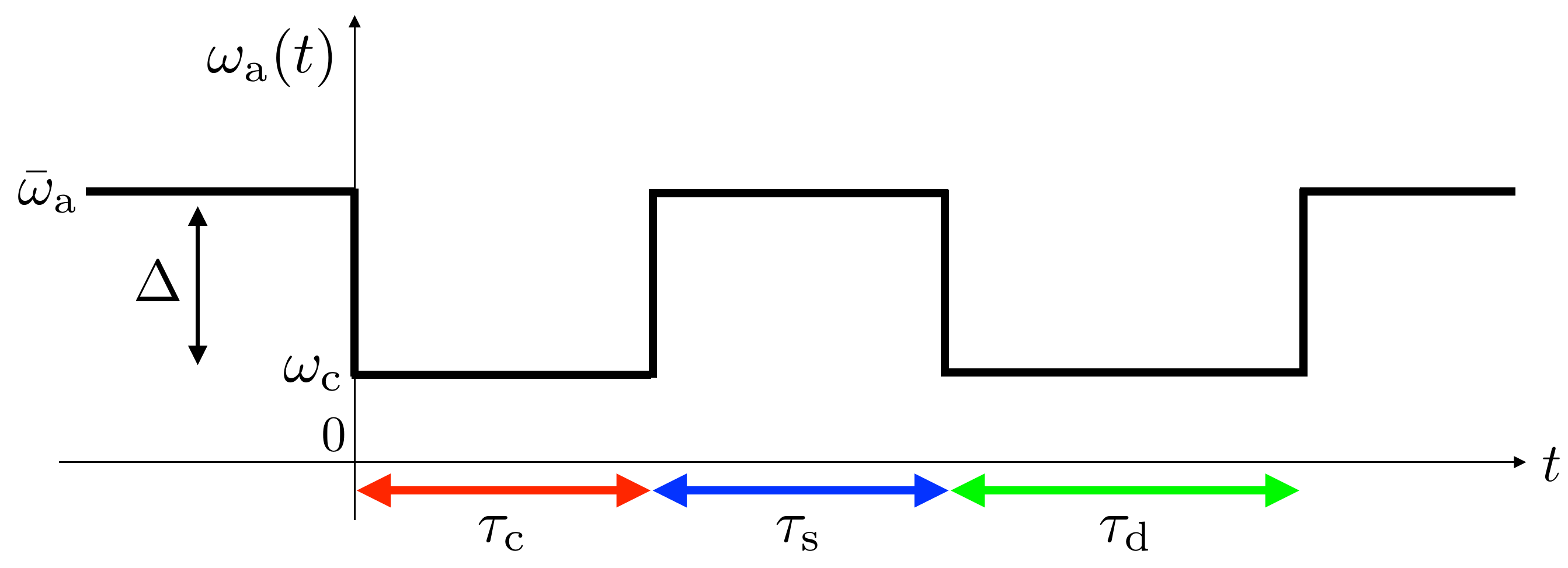}
\caption{(Color online) Alternative charging/discharging protocol. The TLS-cavity coupling $\bar{\lambda}$ stays constant while the TLS energy splitting $\hbar\omega_{\rm a}(t)$ is varied in time in a stepwise manner, therefore affecting the detuning.\label{fig1_supp}}
\end{figure}
The charging/discharging protocol described in the main text relies on turning on and off the interaction between the array of TLSs and the cavity (see Fig.~\ref{fig1} of the main text). Implementing such operation may be experimentally challenging. A more practical way to implement an equivalent charging/discharging protocol can be realized by manipulating the detuning between the TLSs and the cavity mode. When the two subsystems are far detuned, their interaction is effectively quenched.
While the cavity frequency $\omega_{\mathrm{c}}$ is typically determined by fixed geometric features, it is often possible to temporally tune the TLS level splitting  $\hbar\omega_{\mathrm{a}}$ e.g.~by means of an applied magnetic field. 

For a time dependent $\omega_{\mathrm{a}}(t)$ and a fixed interaction strength $\bar\lambda$, the Hamiltonian of the system reads:
\begin{align}\label{eq:Dicke2}
\hat{\mathcal{H}}^{(N)}_{\omega_{\mathrm{a}}(t)}=& \hbar \omega_{\rm c } \hat{a}^{\dagger} \hat{a}+ \omega_{\rm a}(t) \hat{J}_{z} + 2\omega_{\rm c} \bar\lambda \hat{J}_x \left(\hat{a}^{\dagger}+\hat{a}\right)~.
\end{align}
This alternative charging/discharging protocol is sketched in Fig.~\ref{fig1_supp}.

Note that the new Hamiltonian (\ref{eq:Dicke2}) coincides with the Hamiltonian (\ref{eq:Dicke}) considered in the main text during the charging and discharging steps, i.e.~for  $t \in [0,\tau_{\rm c}]\cup [\tau_{\rm c} + \tau_{\rm s}, \tau_{\rm c} + \tau_{\rm s} +\tau_{\rm d}]$. Accordingly, during these time intervals, the dynamics does not change with respect to that discussed in the main text.

For all other times, i.e.~for  $t \notin [0,\tau_{\rm c}]\cup [\tau_{\rm c} + \tau_{\rm s},\tau_{\rm c} + \tau_{\rm s} + \tau_{\rm d}]$, the TLS and the cavity are far detuned, $|\bar{\omega}_{\mathrm{a}}-\omega_{\mathrm{c}}|\gg \omega_{\rm c} \bar\lambda$. In the case of a Rabi QB, the dynamics of each individual TLS+cavity system is well described in the far-detuned limit by the so-called dispersive Hamiltonian~\cite{Schleich_Book, blais_pra_2004}
\begin{align}\label{H_eff_Rabi}
\hat{\mathcal{H}}_{\mathrm{R}}&\approx \hbar \omega_{\mathrm{c}}\hat{a}^{\dagger}\hat{a}+\frac{\hbar \bar\omega_{\mathrm{a}}}{2} \hat{\sigma}_{z} +\hbar g^{2} \Bigg[\left(\frac{1}{\Delta}+\frac{1}{\Sigma} \right) \hat{a}^{\dagger} \hat{a} \hat{\sigma}_{z} \nonumber\\
&+\frac{1}{2} \left( \frac{1}{\Delta}-\frac{1}{\Sigma}\right)+\frac{1}{2} \left(\frac{1}{\Delta}+\frac{1}{\Sigma}\right)\hat{\sigma}_{z}\Bigg]~,
\end{align}
where $\Delta\equiv \bar \omega_{\mathrm{a}}-\omega_{\mathrm{c}}$, $\Sigma\equiv \bar\omega_{\mathrm{a}}+\omega_{\mathrm{c}}$ and $g\equiv \omega_{\rm c}\bar \lambda$. 
Similarly, in the case of a Dicke QB and for $t \notin [0,\tau_{\rm c}]\cup [\tau_{\rm c} +\tau_{\rm s}, \tau_{\rm c} + \tau_{\rm s} + \tau_{\rm d}]$, the dynamics is well described by~\cite{Klimov_Book}
\begin{align}
\label{H_eff}
\hat{\mathcal{H}}^{(N)}_{\bar\omega_{\mathrm{a}}}\approx \hbar \omega_{\mathrm{c}}\hat{a}^{\dagger}\hat{a}+\bar\omega_{\mathrm{a}} \hat{J}_{z}+\delta\hat{\mathcal{H}}~,
\end{align}
where
\begin{align}
\delta\hat{\mathcal{H}}=g^{2} \Bigg[&\left(\frac{2}{\Delta}+\frac{2}{\Sigma} \right) \hat{a}^{\dagger} \hat{a} \hat{J}_{z} +\frac{\hbar}{2} \left( \frac{1}{\Delta}-\frac{1}{\Sigma}\right) \nonumber\\
&+\left(\frac{1}{\Delta}+\frac{1}{\Sigma}\right)\hat{J}_{z}\Bigg]~.
\end{align}
Notice that the Hamiltonian (\ref{H_eff_Rabi}) commutes with both $\hat{\sigma}_{z}$ and $\hat{a}^{\dagger}\hat{a}$ and that, similarly, the Hamiltonian (\ref{H_eff}) commutes with both $\hat{J}_{z}$ and $\hat{a}^{\dagger}\hat{a}$. That is, the interaction is effectively quenched in the far detuned case. Accordingly, no exchange of quanta between the TLS and the cavity mode is allowed when the TLSs are far detuned from the cavity. It follows that the overall dynamics of the TLS+cavity system is the same as the one discussed in the main text. However, we emphasize that the energy expectation can be different in the two cases, when considering the non-interacting stages, i.e.~at $t \notin [0,\tau_{\rm c}]\cup [\tau_{\rm c}+\tau_{\rm s},\tau_{\rm c}+\tau_{\rm s}+\tau_{\rm d}]$. Indeed, there can be considerable differences in the values of the energy injection $\delta E^{\rm on/\rm off}$ at each of the turning-on/off point of the two protocols. To see this, consider for simplicity the turning-on point at time $t=0$ in the limit $g^2N/\Delta,g^2N/\Sigma \ll \min[\bar\omega_{\rm a},\omega_{\rm c}]$. In this case, the term proportional to $g^2$ in Eq.~(\ref{H_eff}) can be neglected and the energy injection amounts to the large value $\delta E^{\rm on}\simeq \hbar(\omega_{\rm c}-\bar{\omega}_{\rm a})N/2$ (see Fig.~\ref{fig1_supp}). On the other hand, $\delta E^{\rm on}$ vanishes for the protocol discussed in the main text, see Fig.~\ref{fig1}(c).

\section{Universality}
\label{app:universal_scaling}
\begin{figure}[t] 
  \centering
\begin{overpic}[width=8.6cm]{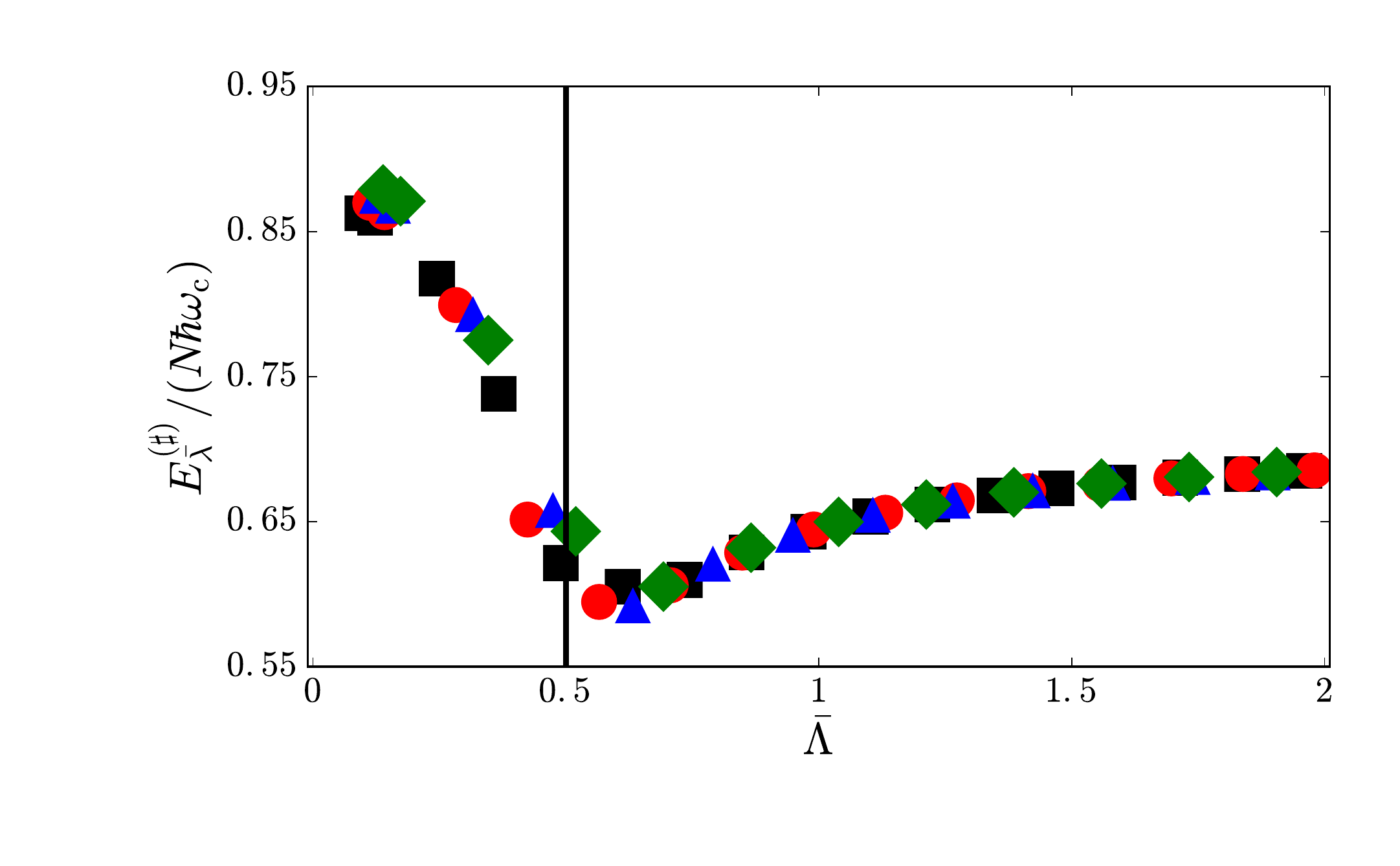}
\put(10,58){(a)}
\end{overpic}
\centering
\begin{overpic}[width=8.6cm]{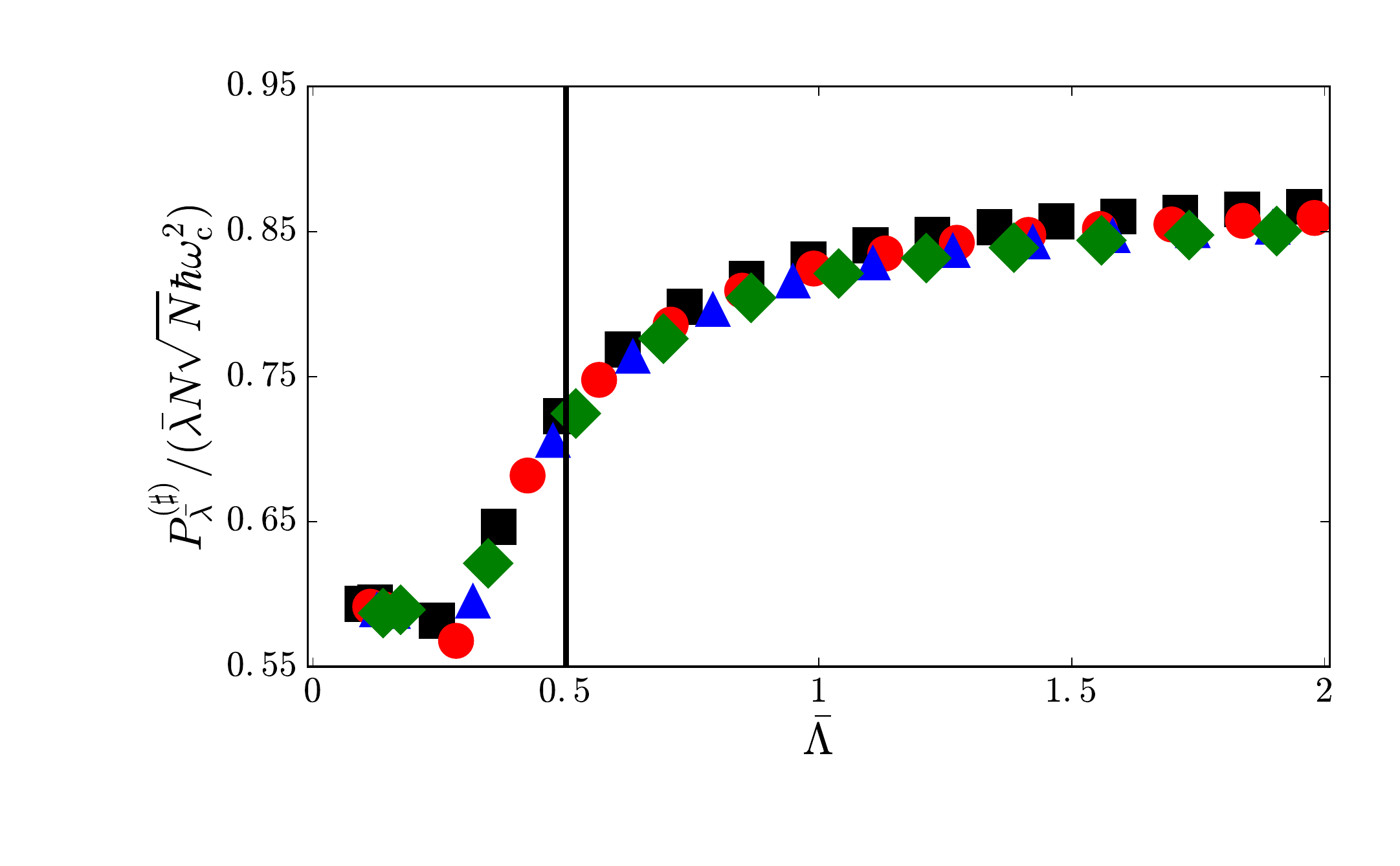}
\put(10,58){(b)}
\end{overpic}
\caption{
(Color online) Panel (a) The maximum stored energy $E^{(\sharp)}_{\bar{\lambda}}$ (in units of $N\hbar \omega_{\rm c}$) is plotted as a function of $\bar{\Lambda}$ for $N=6$ (black squares), $N=8$ (red circles), $N=10$ (blue triangles), and $N=12$ (green diamonds).  Panel (b) The maximum charging power $P^{(\sharp)}_{\bar{\lambda}}$ (in units of $\bar{\lambda} N \sqrt{N}\hbar \omega^{2}_{\rm c}$) is plotted as a function of $\bar{\Lambda}$ for $N=6$ (black squares), $N=8$ (red circles), $N=10$ (blue triangles) and $N=12$ (green diamonds). Vertical black lines indicate the critical coupling $\bar{\Lambda}_{\rm c}=0.5$ at which a superradiant quantum phase transition occurs in the ground state of the system, in the thermodynamic limit~\cite{Emary03}.\label{fig2_supp}}
\end{figure}

Fig.~\ref{fig2_supp}(a) and (b) show the maximum stored energy $E^{(\sharp)}_{\bar{\lambda}}$ and maximum charging power $P^{(\sharp)}_{\bar{\lambda}}$ plotted as functions of the quantity $\bar\Lambda = {\bar \lambda}\sqrt{N}$, which was already introduced in the main text, for various values of $N$. It is remarkable to note that the results for different values of $N$, which were reported in Fig.~\ref{fig3}(b) and (d) of the main text, collapse onto universal curves in both cases.
\section{On the role of a quadratic term in the photon field}
\label{app:Asquared_term}

In certain solid-state experimental implementations of Rabi and Dicke Hamiltonians, it may be necessary to include an extra term proportional to $(\hat{a}^{\dagger}+\hat{a})^{2}$ in the Hamiltonian (\ref{eq:Dicke}).
As mentioned in the main text~\cite{footnote:Asquared_term}, the inclusion of such a term may have important consequences in regard to the achievability of the superradiant quantum phase transition in the thermodynamic limit~\cite{Chen07, Nataf10, Nataf10b, Viehmann11,chirolli_prl_2012,Pellegrino14}. However, as detailed below for the case of finite $N$ considered in this work, the new Hamiltonian with the extra term $\propto (\hat{a}^{\dagger}+\hat{a})^{2}$  takes on exactly the same form of Eq.~(\ref{eq:Dicke}) when re-expressed in terms of properly squeezed bosonic operators $b$, $b^\dagger$. In this case, a renormalized cavity frequency and TLS-cavity coupling constant appear, as we proceed to demonstrate. Consider the new Hamiltonian
\begin{align}\label{eq:full_Hamiltonian_Asquared}
\hat{\mathcal{H}}^{'(N)}_{\lambda_t}&\equiv \hat{\mathcal{H}}^{(N)}_{\lambda_{t}}+\hbar \omega_{\rm c} \kappa (\hat{a}^{\dagger}+\hat{a})^{2} = \hbar \omega_{\rm c } \hat{a}^{\dagger} \hat{a} + \omega_{\rm a} \hat{J}_{z} \nonumber\\
&+ 2 \omega_{\rm c} \lambda_t  J_x \left(\hat{a}^{\dagger}+\hat{a}\right)+ \hbar \omega_{\rm c}\kappa (\hat{a}^{\dagger}+\hat{a})^{2}~.
\end{align}
Introducing the squeezed operator
\begin{align}
\hat{b} = \frac{(\alpha-1)}{2\sqrt{\alpha}} \hat{a}^\dagger + \frac{(\alpha+1)}{2\sqrt{\alpha}} \hat{a}
\end{align}
with $\alpha = \sqrt{4\kappa+1}$, we obtain
\begin{align}
\hat{\mathcal{H}}^{'(N)}_{\lambda_t} = \hbar \alpha \omega_{\rm c } \hat{b}^{\dagger} \hat{b}+ \omega_{\rm a} \hat{J}_{z} + 2 \omega_{\rm c} \alpha^{-1/2} \lambda_t  \hat{J}_x (\hat{b}^{\dagger}+\hat{b})~,
\end{align}
where an irrelevant constant term has been dropped. We therefore immediately see that the new Hamiltonian $\hat{\mathcal{H}}^{'(N)}_{\lambda_t}$ takes the same form as in Eq.~(\ref{eq:Dicke}), but with rescaled parameters. We note, however, that microscopic considerations~\cite{Nataf10b,Viehmann11} reveal that there exist a functional relation between $\kappa$, $\bar{\lambda}$ and $N$,
\begin{equation}\label{eq:xi}
\kappa= \xi N \bar{\lambda}^{2}~,
\end{equation}
where $\xi$ is a free parameter \cite{Nataf10b, Viehmann11}. Consequently, when the $(a^\dagger+a)^2$ term is accounted for, a change in the coupling $\bar{\lambda}$ would be accompanied by a change in $\kappa$, leading to both a change in the dressed cavity mode frequency $\omega'_{\rm c}=\alpha\omega_{\rm c}$ and in the cavity eigenstates. This unfortunate situation does not occur when the more realistic protocol described in Appendix ~\ref{AppA} is used. In this case, the coupling stays fixed and only the TLS frequency changes. The new Hamiltonian remains therefore formally the same as in the main text as well as the initial state (involving now the $N$-photon Fock state associated to the operator $b$, instead of $a$), leaving unchanged the analysis described in the main text.


\begin{thebibliography}{99}
\bibitem{Prefactors_1}
The prefactors can be understood considering the fact that $y=1-\cos{x}$ has absolute maximum $y_{\max}=2$ at $x_{\max}=\pi$ ($\mathrm{mod}\left[2\pi\right]$), while $y=(1-\cos{x})/x$ has the absolute maximum $y_{\max}\approx 0.724$ at $x_{\max}\approx 2.331$ as can be easily achieved by numerical investigation.
\end{thebibliography}
\end{document}